\documentclass[a4paper,12pt]{article}

\pdfoutput=1
\usepackage[margin=1.0in]{geometry}

\usepackage{graphics}
\usepackage{graphicx}
\usepackage{epsfig}
\usepackage{subfigure}

\usepackage{amssymb}

\usepackage[bookmarksopen=false]{hyperref}

\begin{document}




\title{Study of cluster shapes in the MIMOSA-5 pixel detector}




\author{
{\L}. M\c{a}czewski$^1$, M.Adamus$^2$, J.Ciborowski$^1$, \\
G. Grzelak$^1$, P. {\L}u\.zniak$^3$ P.Niezurawski$^1$, A.F.\.Zarnecki$^1$\\
\vspace{.3cm}\\
%
1- Institute of Experimental Physics, University of Warsaw \\
Ho\.za 69, 00-681 Warszawa, Poland
\vspace{.1cm}\\
2- The Andrzej Soltan Institute for Nuclear Studies, \\
Ho\.za 69, 00-681 Warszawa, Poland
\vspace{.1cm}\\
3- University of {\L}\'od\'z, Faculty of Physics and Applied Informatics, \\
Pomorska 149/153, PL-90236 {\L}\'od\'z, Poland
}

\maketitle

\begin{abstract}
Beamstrahlung will constitute an important source of background in a pixel vertex detector at the future International Linear Collider. Electron and positron tracks of this origin impact the pixel planes at angles generally larger than those of secondary hadrons and the corresponding clusters are elongated.
We report studies of cluster characteristics using test beam electron tracks incident at various angles at a Mimosa-5 pixel matrix.
\end{abstract}



\section{Introduction}
\label{intro}

Precise spatial reconstruction of tracks and secondary vertices is imperative for flavour identification in experiments at the future International Linear Collider, where electron and positron beams will be colliding at the centre-of-mass energies up to 1~TeV \cite{Accelerator}. This can be achieved by means of a multi-layer pixel vertex detector, enveloping the beam pipe near the interaction point. Such a detector may be built of rectangular pixel matrices, arranged quasi cylindrically (with possible end caps) \cite{Detector}. Charged final state particles from the $e^+e^-$ interactions will traverse the detector layers and  each particle will deposit charge in a group of pixels, forming a cluster. The vertex detector will be also exposed to an important source of background - beam\-strahlung. This is an electromagnetic process in which a brems\-strahlung photon creates an electron - positron pair in the interaction region of the colliding beams. Subsequently these pairs will enter the vertex detector among other products of the interaction and create numerous additional clusters.

\begin{figure}[!htbp]
	\begin{center}
		\resizebox{9.0cm}{!}{
			\includegraphics[]{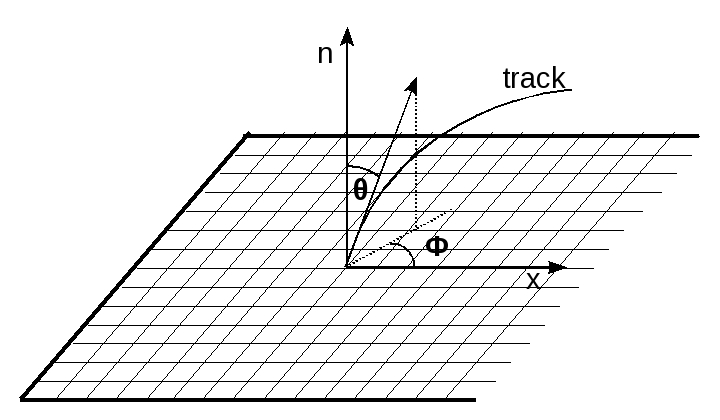}
		}
		\caption[]{Definition of the polar and azimuthal angles, $\theta$ and $\phi$, in the MIMOSA-5 coordinate system with the $X$ axis parallel to the MIMOSA-5 edge; n - normal to the pixel plane.}
		\label{LuPawel1}
	\end{center}
\end{figure}

The orientation of a track w.r.t. a given pixel matrix plane may be described in terms of two angles as shown in fig.~\ref{LuPawel1}: ($i$) the polar angle $\theta$ which is the angle between the track and the normal to the matrix plane and ($ii$) the azimuthal angle $\phi$ which is the angle between the direction of the track projected to the matrix plane and one of the matrix axes.

Simulations have shown~\cite{PLuzniak}: ($i$) the presence of the beamstrahlung background deteriorates flavour identification; ($ii$) the angles $\theta$ and $\phi$ of beamstrahlung tracks differ from those of the final state particles from $e^{+}e^{-}$ interactions. Meanwhile one expects that the cluster shape depends primarily on the angle of incidence, $\theta$, of a given track. Since the beamstrahlung tracks traverse pixel matrices predominantly at large angles $\theta$, one may expect that the corresponding clusters will be elongated for sufficiently large values of this angle. The aim of the present paper is to study this effect, which might appear helpful for identification of beamstrahlung clusters on the basis of their shape and orientation on the pixel plane.

Monolithic Active Pixel Sensor, MAPS, is one of the vertex detector technology concepts presently under consideration. Studies show that the MAPS matrices may have spatial resolution of the order a few $\mu$m, almost 100\% detection efficiency, possess good resistance to radiation and can be thinned down to about $50$ $\mu$m \cite{Deptuch1}.
We have used the  MIMOSA-5\footnote{\textbf{M}inimum \textbf{I}onizing particle \textbf{MOS} \textbf{A}ctive pixel sensor} MAPS prototype for our  studies. The device was fabricated in the AMS $0.6~\mu$m CMOS technology. The epitaxial layer, which constitutes the sensitive volume of the detector, was $14~\mu$m thick. We have used a module of $510 \times 512$ pixels ($8.7 \times 8.7$ mm$^{2}$). Pixel dimensions were $17 \times 17~\mu$m$^{2}$ and those of the charge collecting diodes $4.9 \times 4.9~\mu$m$^{2}$. The data acquisition sytem is described elsewhere \cite{VME-DAQ}.

\section{Experimental setup}
\label{exp}

\begin{figure}[!htbp]
	\begin{center}
		\resizebox{12.0cm}{!}{
			\includegraphics[]{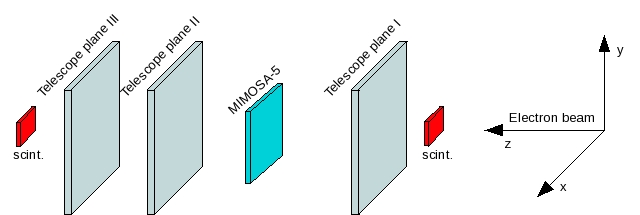}
		}
		\caption[]{Schematic view of the experimental setup (see text).}
		\label{CoordinatesSystem}
	\end{center}
\end{figure}

The experimental setup included the MIMOSA-5 matrix, the beam telescope and trigger counters, as shown in fig.~\ref{CoordinatesSystem}. The MIMOSA-5 matrix was mounted on an adjustable mechanical support, enabling rotations around the $X$ and $Y$ axes and was contained in a polystyrene box for thermal insulation. The matrix was oriented  manually to the desired angles before each data taking run using the angular scale of an accuracy of approximately $\pm 1^\circ$. Precise determination of the angles was performed offline in the alignment procedure (see below).

The beam telescope consisted of a set of three microstrip units, each unit built of two planes of perpendicularly oriented microstrips. One unit was placed in front and two other behind the MIMOSA-5. Each telescope plane had an area of $32\times32$ mm$^{2}$ and the overall detection efficiency higher than 99\% \cite{DevisBeam}. The beam telescope enabled reconstruction of the beam track  position at the MIMOSA-5 matrix with accuracy  better than $10$ $\mu$m at the offline analysis stage. Three small plastic scintillator detectors were placed on the beam (two in front and one behind the beam telescope), serving as trigger counters.\\
The MIMOSA-5 matrix was exposed to the DESY electron beam of energies from 1 to 6~GeV. The measurements were done for various orientations of the matrix w.r.t. the beam. Approximately 5k triggers were collected for each run.

\section{Measurement of cluster shapes}
\label{clshape}

\begin{figure}[!htbp] 
	\begin{center}
	\subfigure[]{
		\includegraphics[width= 7.0cm,angle=90]{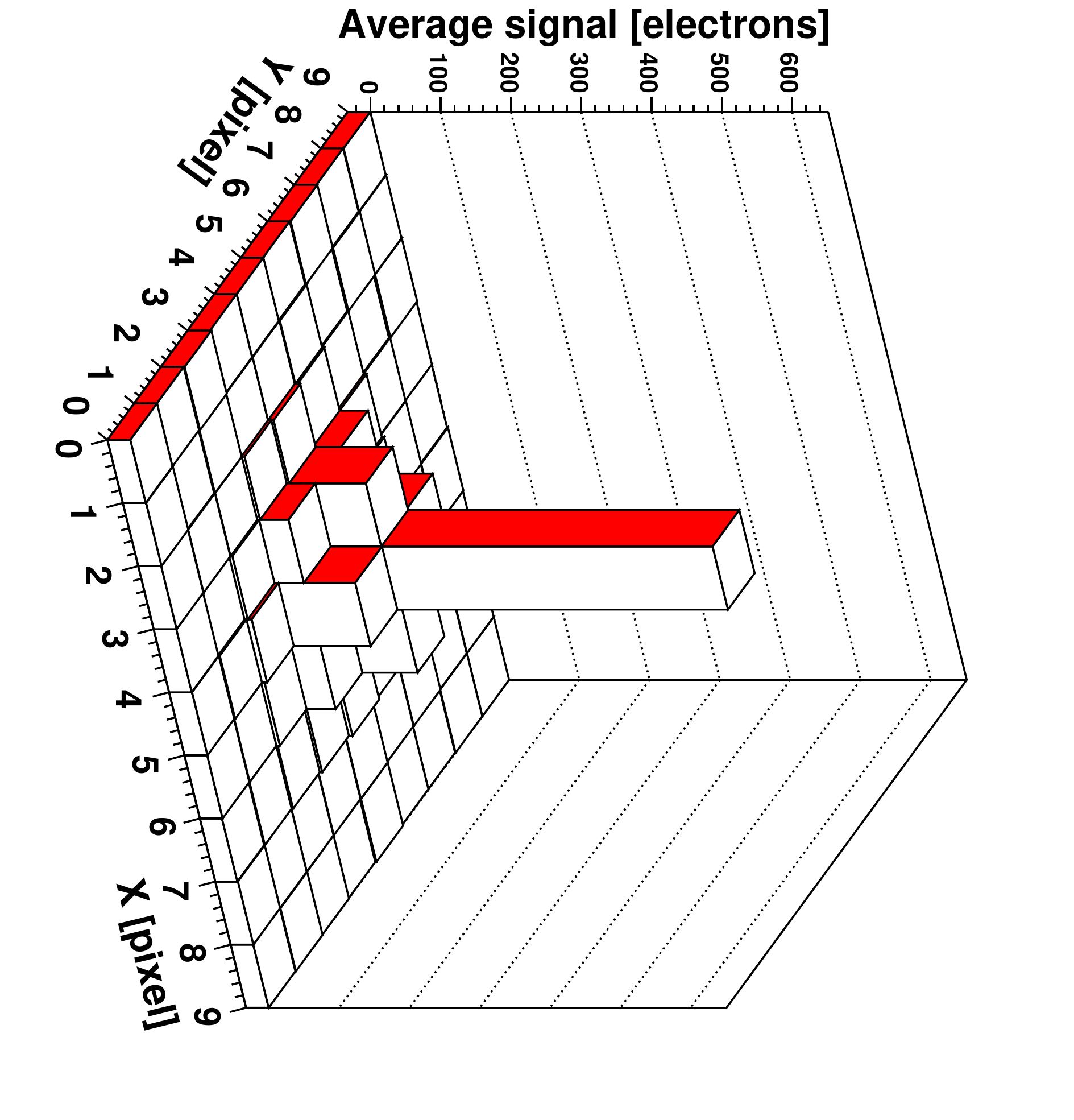}
		\label{MeanCl2012}
	}
	\hspace{0.5cm}
	\subfigure[]{
		\includegraphics[width= 7.0cm,angle=90]{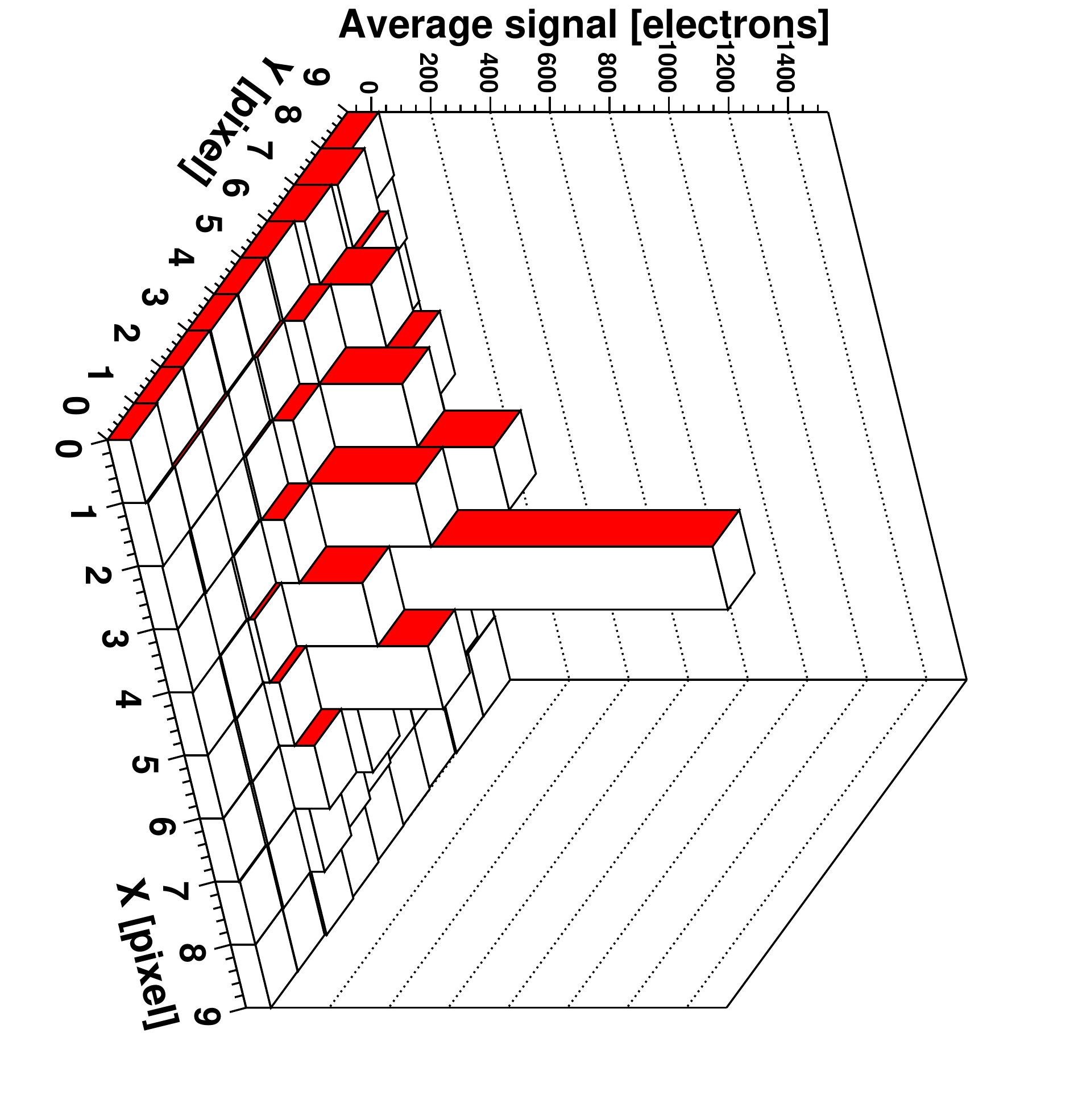}
		\label{MeanCl2021}
	}
	\caption{Spatial distribution of charge in an averaged cluster for two different incident angles: (a) $\theta = 4.3^{\circ}$ and $\phi = -44.0^{\circ}$, (b) $\theta = 78.1^{\circ}$ and $\phi = -37.8^{\circ}$.}
	\label{MeanCl}
	\end{center}
\end{figure}

The first step in the data analysis was the evaluation of the 
pedestals and noise levels  for each pixel. Clusters of pixels were reconstructed as follows. Pixels with locally maximal charge (the seed) were identified as the highest among 8 adjacent pixels. It was required that the signal to noise ratio, $S/N$, for the seed  was greater than~5. In order to remove fake clusters (formed around malfunctioning pixels) an additional cut was applied. For this purpose the following variables were defined: $(i)$ $\mathcal{S}_{8} = \sum_{i = 1}^{8}q_{i}$ - the total charge collected in 8 pixels adjacent to the seed and $(ii)$ $\mathcal{N}_{8}=\sqrt{\sum_{i=1}^{8}N_{i}^{2}}$ - average noise of these pixels, where $q_{i}$ and $n_{i}$ are the charge and the noise of the $i$-th pixel, respectively. If the condition $\mathcal{S}_{8}/\mathcal{N}_{8} > 0.5$ was fulfilled, the group of $9\times9$ pixels around the seed was taken under consideration. The 81 pixels were sorted in the descending order with respect to the charge. It was then accepted that a cluster is formed of the first $N_c$ consecutive pixels, counting from the seed (highest charge), which carried $90\%$ of the total charge of the $9\times9$ group. Average values of the number of pixels forming a cluster, $N_c$,  ranged from about 6 to 14, for almost perpendicular ($\theta \approx 4^{\circ}$) and very inclined tracks ($\theta \approx 78^{\circ}$), respectively. It was assumed that the charge-weighted centre of gravity of a cluster coincided with the track position at the matrix surface. Cluster positions and track positions extrapolated from the telescope constituted partial input for the offline alignment. This procedure was based on $\chi^{2}$ minimisation involving parameters describing the position and angular orientation of the matrix w.r.t. the telescope reference frame. The procedure allowed computing precise values of the $\theta$ and $\phi$  angles for a given, manually preset, orientation of the pixel matrix. The accuracy of the angle determination using alignment was better than $0.1^{\circ}$.\\
Particles passing through the detector at low incident angles leave statistically round clusters.
Since in a MAPS detector the charge transport is by diffusion, it is expected that clusters arising from sufficiently inclined tracks should be elongated in the track direction projected to the pixel plane while unaltered in the perpendicular direction. We have measured the magnitude of elongation as a function of the incident angle $\theta$ and determined orientation of elongated clusters on the matrix plane in terms of the angle $\phi$.\\
Shapes of individual clusters are subject to statistical fluctuations due to fluctuations of charge deposited in pixels. For the purpose of visualisation we averaged clusters over approx. 1.5~k events taken at the same $\theta$ and $\phi$ settings. In fig.~\ref{MeanCl} we  show  two distinct cases: averaged clusters arising from ($a$) electron tracks incident at $\theta \approx 4^\circ$ and ($b$) tracks incident at $\theta\approx 78^\circ$; the elongation of the averaged cluster is clearly visible in the latter case.

The shape of a single cluster, its longitudinal and transverse dimensions, may be obtained from the following procedure. The charge distribution matrix in a given cluster was defined as:

\begin{equation}
\left(\begin{array}{cc}
	\sum_{i = 1}^{N_{c}}\frac{q_{i}}{Q}\left(x_{i}-\overline{x}\right)^{2} & \sum_{i = 1}^{N_{c}}\frac{q_{i}}{Q}\left(x_{i}-\overline{x}\right)\left(y_{i}-\overline{y}\right)\\
	\sum_{i = 1}^{N_{c}}\frac{q_{i}}{Q}\left(x_{i}-\overline{x}\right)\left(y_{i}-\overline{y}\right) & \sum_{i = 1}^{N_{c}}\frac{q_{i}}{Q}\left(y_{i}-\overline{y}\right)^{2} \\
	\end{array} \right),
\label{Matrix}
\end{equation}

where $Q$ is the cluster charge, $q_{i}$ is a charge of the $i$-th pixel, $x_{i}$, $y_{i}$ are its positions in the MIMOSA-5 coordinates sytem and the $\overline{x}$ and $\overline{y}$ are the coordinates of the considered cluster in the latter coordinates system. The $\overline{x}$ and $\overline{y}$ were assumed to be the charge-weighted centre of gravity of the cluster:

\begin{equation}
\overline{x} = \sum_{i = 1}^{N_{c}}\frac{q_{i}}{Q} x_{i},~~~
\overline{y} = \sum_{i = 1}^{N_{c}}\frac{q_{i}}{Q} y_{i}.
\label{CentreOfGravity}
\end{equation}

\begin{figure}[!htbp] 
	\begin{center}
	\subfigure[]{
		\includegraphics[width=7.0cm,angle=90]{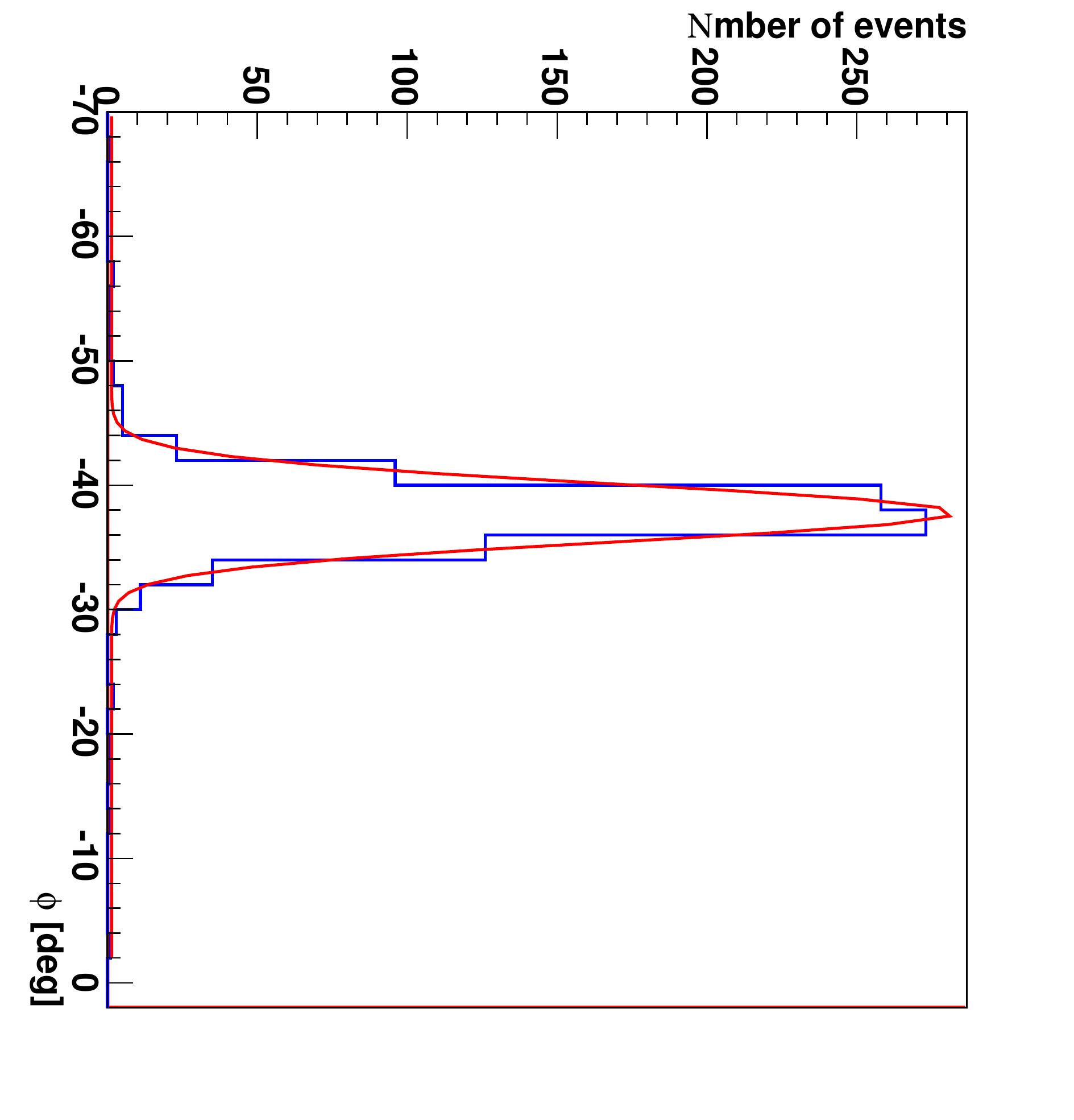}
		\label{PhiReco1}
	}
	\hspace{0.5cm}
	\subfigure[]{
		\includegraphics[width=7.0cm,angle=90]{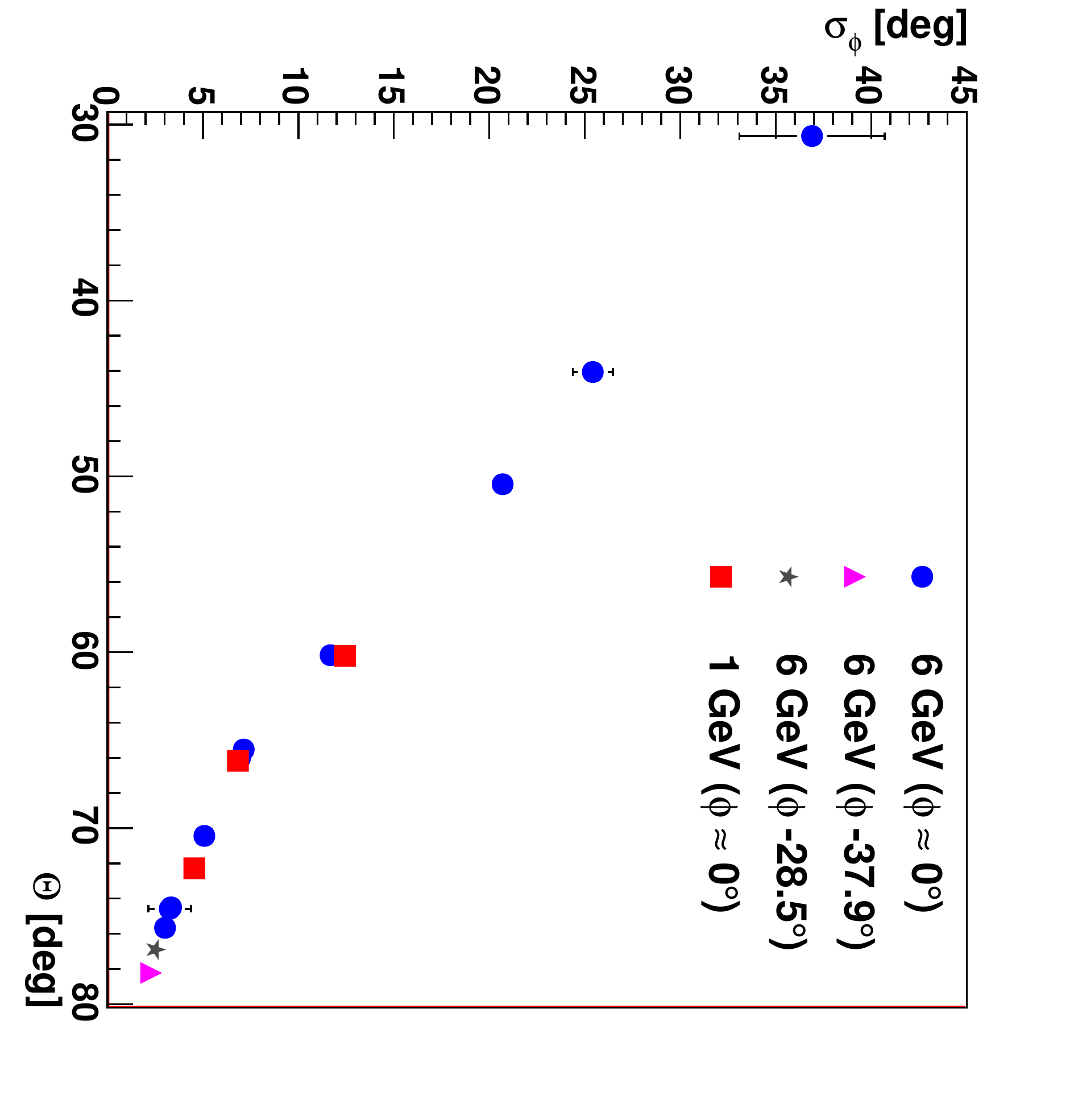}
		\label{PhiReco2}
	}
	\caption{(a) Distribution of the reconstructed $\phi$ angle, (b) dispersion of the $\phi$ distribution as a function of the incident angle $\theta$.}
	\label{PhiReco}
	\end{center}
\end{figure}

Diagonalisation of the matrix~(\ref{Matrix}) allowed to determine the eigenvectors, $\vec v_{L}$ and $\vec v_{T}$, which coincide with the longitudinal and transverse axes of the cluster, respectively, provided the cluster is sufficiently elongated. The corresponding eigenvalues, $\sqrt{\lambda_{L}}$ and $\sqrt{\lambda_{T}}$, are proportional to the cluster longitudinal and transverse dimensions.\\
Given the eigenvectors $\vec v_{L}$ and $\vec v_{T}$, one may evaluate the azimuthal angle of the cluster w.r.t. the axis of the pixel matrix, $\phi_c$, for each individual cluster. The distribution of this angle for a selected data sample of 1~k clusters, taken with the beam setting of $\phi = -37.8^{\circ}$ and $\theta = 78.1^{\circ}$, is presented in fig.~\ref{PhiReco1}.

The peak position of the distribution in that sample is $\phi_c = -37.9^{\circ}\pm0.1^{\circ}$, in excellent  agreement with the actual value quoted above as obtained from alignment. The dispersion of this distribution, $\sigma_{\phi_{c}}$, obtained from fitting the Gaussian, is $2.4^{\circ}\pm0.1^{\circ}$.\\
The accuracy of determining the axes of a cluster diminishes as the track becomes steeper and clusters less elongated, which is reflected in increasing dispersion of the $\phi_{c}$ distribution. Measurements of the dispersion $\sigma_{\phi_{c}}$ for several values of $\theta$ are shown in fig.~\ref{PhiReco2}, performed using 1 and 6 GeV electron beams and for various $\phi$ values.

\begin{figure}[!htbp]
	\begin{center}
		\resizebox{8.0cm}{!}{
			\includegraphics[angle=90]{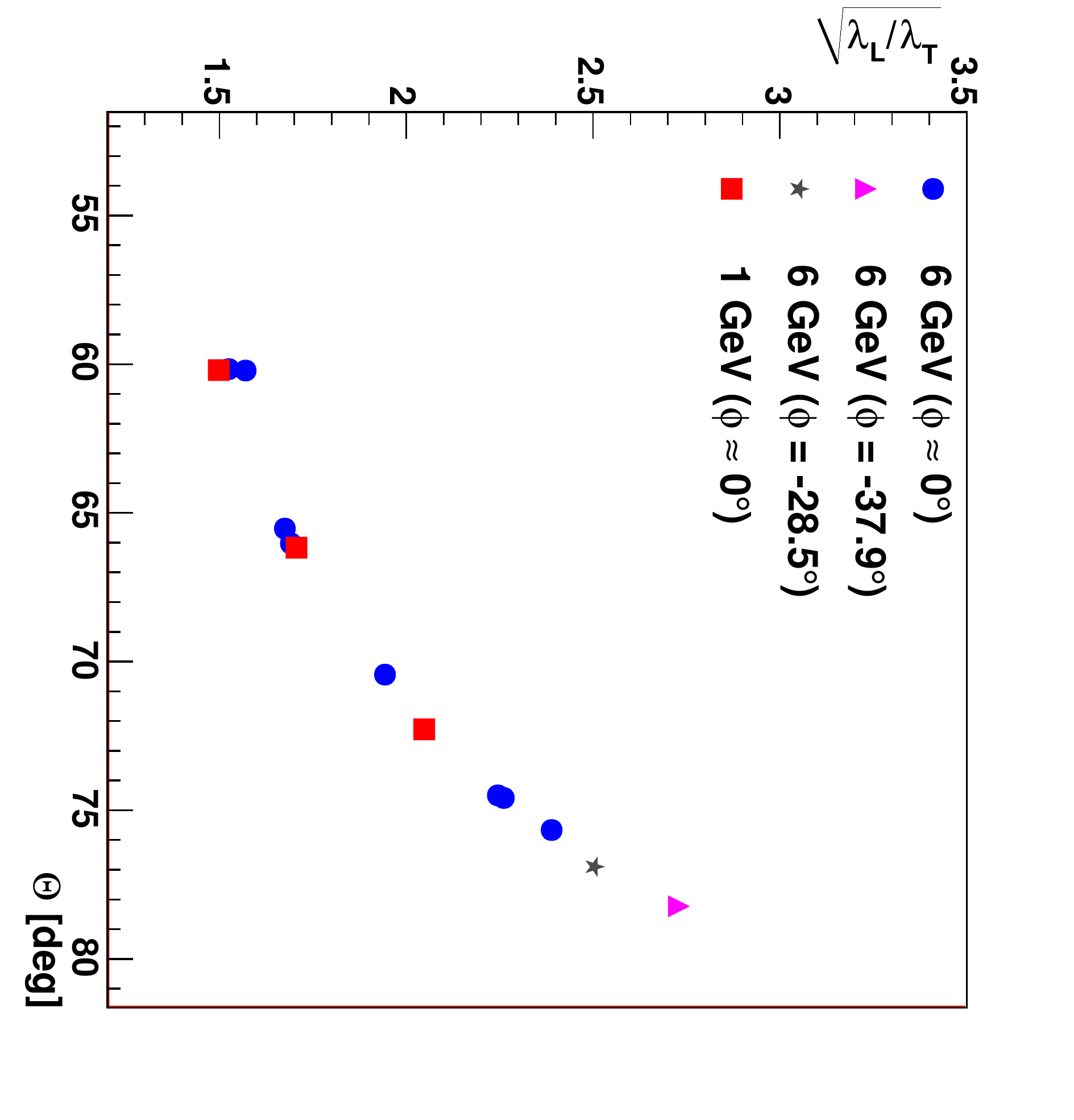}
		}
		\caption[]{Measured ratio of the longitudinal and transverse dimensions of a cluster as a function of the $\theta$ angle.}
		\label{Eigen1ToEigen2}
	\end{center}
\end{figure}

The ratio $\sqrt{\lambda_{L}/\lambda_{T}}$ is a measure of cluster elongation. Measured dependence of this quantity on the incident angle $\theta$ is shown in fig. \ref{Eigen1ToEigen2}. Thus we conclude that the cluster elongation depends strongly on $\theta$ (while there is little or no dependence on beam energy in the studied range). We estimate that this observation is valid for $\theta > 60^{\circ}$ since for tracks incident at lower angles th elongation of clusters cannot be reliably determined. This can be explained by the fact that in these clusters approximately the same numbers of pixels are involved in charge collection.\\

\section{Conclusions}
\label{conclusions}

We have studied cluster shapes using the MIMOSA-5 MAPS detector exposed to 1 and 6 GeV electron beams. Cluster elongation was measured for various orientations of the pixel matrix w.r.t. the beam axis (angles $\theta$ and $\phi$). The results may be summarised as follows: ($i$) precision of reconstruction of the azimuthal angle $\phi$ increases with increasing incident angle $\theta$; ($ii$) clusters are elongated for  $\theta > 60^{\circ}$ and the effect grows rapidly with increasing $\theta$, independent of the angle $\phi$; ($iii$) no energy dependence of the above effects is observed.

\section{Acknowledgements}
\label{acknowledgements}

We thank the DESY Directorate for the possibility to use the test beam facilities, the IReS (Strasbourg) and in particular Dr~M.~Winter for numerous discussions and lending us the MIMOSA-5 matrix. We also thank Dr~M.~I.~Gregor for her comments on the text. This work was partialy supported from the research funds of the Polish Ministry of Science and Higher Education as a part of the research project. This work was supported by the Commission of the European Communities under the 6$^{th}$ Framework Programme ''Structuring the European Research Area'', contract number
RII3-026126.



\begin{thebibliography}{00}




\bibitem{Accelerator} ILC Reference Design Report Volume 3 - Accelerator, arXiv:0712.2361v1 [physics.acc-ph], 2007.
\bibitem{Detector} ILC Reference Design Report Volume 4 - Detectors, arXiv:0712.2356v1 [physics.ins-det], 2007.
\bibitem{PLuzniak} P. {\L}u\.zniak, in Proceedings of the International Linear Collider Workshop, Hamburg, 2007, edited by S. Riemann, eConf C0705302, Trk07 (2007) (to be published). 
\bibitem{Deptuch1} G. Deptuch et al., Nucl. Instr. and Meth. Phys. Res. A 511 (2003) 240-249.
\bibitem{VME-DAQ} G. Claus et al., IEEE Trans. Nucl. Sci. NS-52 (4) (2005).
\bibitem{DevisBeam} D. Contarato et al., Nucl. Instr. and Meth. Phys. Res. A 565 (2006) 119-125.
\bibitem{MAPS1} R. Turchetta et al. Nucl. Instr. and Meth. Phys. Res. A 458 (2001) 677.
\bibitem{DMeier} D. Meier, Ph.D. Thesis, University of Heidelberg, Germany, 1999.
\bibitem{DeptuchPhD} G. Deptuch, Ph.D. Thesis, Universit\'e Louis Pasteur, Strasbourg (France), 2002.

\end{thebibliography}
\end{document}